**Glass Transition Temperature in Polystyrene Supported Thin Films: a SPM-based Investigation of the Role of Molecular Entanglement**

Franco Dinelli, Tommaso Sgrilli*, Andrea Ricci, Paolo Baschieri (INO, CNR, Pisa, Italy)

Pasqualantonio Pingue (Laboratorio NEST - Scuola Normale Superiore, and Istituto Nanoscience -CNR, Pisa, Italy)

Manjunath Puttaswamy (The Interdisciplinary Nanoscience Centre (iNANO), Faculty of Science, Aarhus University, Aarhus, Denmark)

Peter Kingshott (Industrial Research Institute Swinburne (IRIS), Faculty of Engineering and Industrial Sciences, Swinburne University of Technology, Hawthorn, 3122 VIC, Australia)

**Abstract**

The viscoelastic properties of thin polymeric films represent a central issue, especially for nanotechnological applications. In particular, it is highly relevant the dependence of viscoelasticity on the temperature. For polystyrene it is known that the glass transition temperature is dependent on the film thickness. At present, there is wide agreement on the importance of the two interfaces that the films form with the air and with the substrate. The relevance of molecular entanglement has been also stressed for the case of suspended films. However, the role of molecular entanglement on the glass transition temperature of supported films still remains elusive. In order to investigate the viscoelastic properties of thin films on the nanoscale, we have employed a scanning probe microscope suitably modified in order to monitor the indentation of a tip into a polymeric film during a given lapse of time with the application of a constant load. Thin polystyrene films have been prepared on a range of different substrates: native silicon oxide, hydrogen-terminated silicon and polystyrene brushes. In particular, we have considered polystyrene molecules with molecular weight values below and above the critical value for the occurrence of molecular entanglement. We find that, for samples where molecular entanglement can occur accompanied by a strong interaction with the substrate either by means of chemical bonds or physisorption, the glass transition temperature of thin films increases back to values comparable with those of thick films. This phenomenon is envisioned to be of great relevance in those cases where one needs to improve the adhesion and/or to control the viscoelastic properties of thin films.

**Corresponding author:  franco.dinelli@ino.it**

\*Presently:   Surface Science Research Centre, The University of Liverpool, Liverpool L69 3BX, UK

## 1. Introduction

Polymeric thin films are of high interest for several applications ranging from surface coatings to lithographic processes [1,2] and dielectric layers in organic electronics [3]. Important issues common to most of these applications are represented by the need of improving film adhesion with the substrate and the need to control the film viscoelastic properties as a function of temperature (T). These aspects become more crucial when one employs thinner films. In the microelectronics industry, for instance, dimension dependent properties are expected to pose significant challenges especially when patterning photoresists by advanced lithography below 100 nm. In these systems, processing conditions may need to be optimized as a function of film thickness.

Nowadays it is widely known that confinement effects lead to modifications of molecular dynamics affecting glass transition temperature ($T_g$). The first report of such a phenomenon dates back to the mid-nineties [4] for the case of polystyrene (PS). Since then many investigations have been carried out showing that this phenomenon depends on many parameters such as the material, the substrate employed and the molecular weight ($M_W$) [5-12]. Although one can find some discrepancies in the literature [13], there is however wide agreement on the importance of the two interfaces that the film forms: the one with the air and the one with the substrate.

Concerning the air interface, in a recent paper, we have addressed the issue regarding the drop in $T_g$ at the air interface and the depth of this region [14]. Our data indicate that, if a reduction in $T_g$ occurs, the surface region involved has a thickness inferior to 3 nm. These findings are in agreement with a wide range of previous studies [15-20]. More recently, this behaviour has been specifically confirmed for unentangled PS films by a neutron scattering study [21].

Concerning the role of the interaction with the substrate, many studies have been carried out. In the literature of the last decade, the relevance of molecular entanglement, occurring above a critical value ($M_C \approx 30$ kDa, for PS) [22], has been also stressed. However, up to now, it has been proven to be more relevant for unsupported films compared to supported ones [7,9-10]. For supported films, a $T_g$ enhancement with reducing thickness has been observed for PMMA on the native oxide substrate of Si wafers ($SiO_x$) [5,8], whereas a reduction is generally observed for PS [5-6]. On UV treated silica, both PS and PMMA show a $T_g$ reduction [23]. On hydrogen (H)-terminated silicon surfaces, a $T_g$ reduction is observed for PMMA [8], whereas there is no general agreement for PS. Some report an increase in $T_g$ [24-26], and some a reduction [4,16]. All these data have been obtained for $M_W$ values above $M_C$. A recent study proves that H-terminated silicon surfaces do not oxidize during the process of thermal annealing after film spincoating [27]. A possible explanation of these discrepancies might be represented by the instability of H-terminated silicon surfaces before film spincoating, making the sample preparation crucial [24]. On polymeric substrates the results are also quite controversial. Torkelson and co-workers have observed for PS a $T_g$ increase on P2VP and PMMA, difficult to explain given the immiscibility of the polymers considered [28]. More recently, a very important study of PS deposited on metal surfaces has shown the crucial role of thermal annealing due to the formation of adsorbed layers with a reduced mobility at the metal interface [29]. The issue, regarding a change in molecular mobility at the substrate interface, had been also underlined by other authors previously (see for instance Ref. 12).

In this paper, we report on an investigation of the viscoelastic properties of thin PS films, addressing issues such as the role of the interaction of PS molecules with the substrate and the role of molecular entanglement. Previously, viscoelastic properties of films have been measured employing various techniques [30-31]. In our case, we have developed a setup based on an atomic force microscope and a custom-developed lithographic software capable of controlling the relative position of the tip with respect to the sample surface while fixing parameters such as applied load, temperature and time of contact [32]. The tip can be placed in contact with a surface at a given load and temperature while monitoring the tip penetration and the cantilever torsion during a given time interval (Fig. 1). Whereas quantitative data on the film viscosity is beyond reach due to the complexity in modeling the tip/surface system, via measuring the occurrence of viscous yield as a function of temperature one can infer the $T_g$ value of the film under investigation [14].

We have restricted our attention to films of thickness values low enough so that the $T_g$ is below the bulk value ($T_g^{bulk}$) when the films are spin-coated on a non-interacting surface (in our case $SiO_x$). Films of the same thickness have been placed on substrates such as H-terminated silicon wafers and PS brushes. The brushes have been obtained by grafting PS molecules onto $SiO_x$. Both films and brushes have been made with PS molecules of $M_W$ values below and above $M_C$.

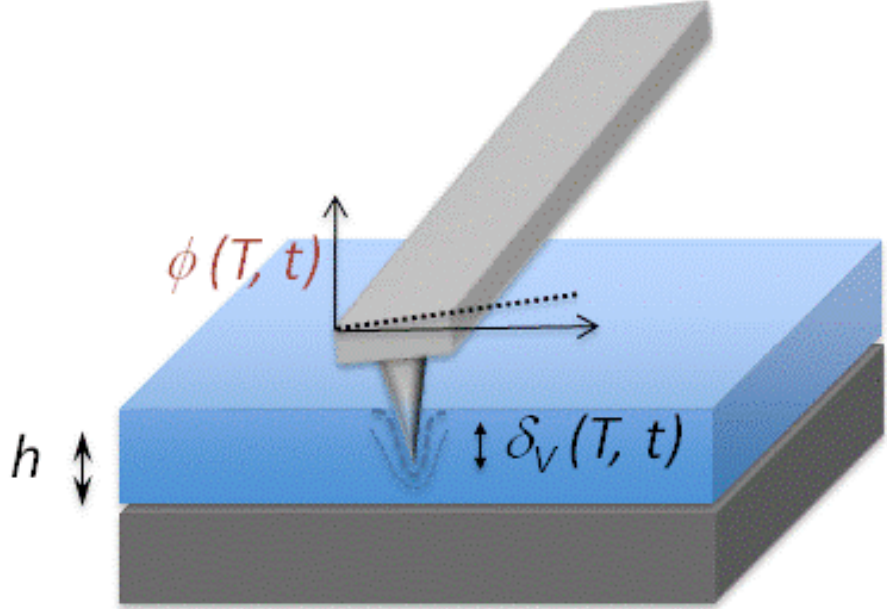

**Fig. 1:** Schematic drawing of an indentation experiment. The tip is placed in contact with a polymer film of a given thickness ($h$). The cantilever deflection is kept constant while monitoring the tip penetration ($\delta_V$) and the torsion signal ($\phi$) of the cantilever at a given temperature ($T$) and for a given lapse of time ($t$).

## 2. Experimental Section

Our experimental setup consists of a commercial head (NT-MDT, SMENA) with the cantilever moving respect to the sample [32]. The head stands with three micrometric screws on a large metallic basis where a Peltier cell is placed in order to vary the sample temperature. The head is controlled via an open and flexible software that allows the acquisition of the relative distance, normal and lateral force during the 'contact' time. The Peltier cell is driven by a commercial electronic (Thorlabs, TED350). This setup, thanks to the reduced number of elements forming the loop between tip and sample is very stable. The film is thermalized before any measurement (0,5°C/min). The tip is not heated but is kept close to the surface during thermalization, thus it is not at RT as the sample irradiates. The $T_g$ values obtained for thick films are in good agreement with those obtained for the bulk and with other techniques on PS thick films. Thin films should thermalize more quickly than thick ones. For these reasons the T of the film is considered equal to the value obtained with the calibration procedure [14].

As shown in a previous publication [14], $z$ increases immediately after the feedback is switched on and flattens out at a value of around 25 nm, depending on the history of the applied voltage to the piezoactuator. This is a creep effect of the piezoactuator. It can be evaluated on a rigid sample such as a piece of silicon wafer and its contribution then considered in the data analysis: it has the opposite sign from the tip penetration.

Our experimental approach consists in measuring the viscoelastic properties on the nanoscale by means of Scanning Probe Microscope (SPM) operating in a new configuration [32]. At a given T, the tip is brought into contact with the surface (see Fig. 2(a)). As the prefixed deflection value is reached, the feedback circuit is switched on in order to maintain the deflection constant. The lateral force ($\phi$) and the relative position of the tip ($z$) with respect to the surface are then monitored for a certain lapse of time. We find that 60" is the appropriate timescale to correctly observe the occurrence of viscous yield [14]. This operation can be repeated as a function of the applied load ($F_N$) and of the sample T. $F_N$ can be varied in a range of values from 50 to 300 nN.

The cantilevers are rectangular as they can be easily calibrated using Sader's method [33-34]. The elastic constant ($k_C$) is selected in order to allow us to apply a wide range of normal load values, the lower $k_C$ the larger the normal load range [35]. We have found that a $k_C$ value of around 5-10 N/m represents a good compromise [36]. The calibration of the cantilever deflection is performed on a stiff surface (silicon oxide wafer). The torsion constant ($k_{yT}$) cannot be calibrated. According to theoretical formulas and the dimensions of the cantilever employed, we can estimate that $k_{yT} \approx 25$ x $k_C$ [37]. Therefore, small lateral movements can determine quite large lateral forces.

The tip cleaning is performed with piranha solution to remove organic contaminants due to the storing package (made of gel-pack ©) and sometimes to remove the PS molecules attached during film analysis at high T. The piranha treatment produces hydrophilic surfaces. As the PS surfaces are hydrophobic, there is little chance that the molecules attach to the tip. Scanning Electron Microscope (SEM) has been employed to measure the tip radius before and after the measurements. The substrates employed are silicon wafers with native oxide. The silicon wafers are ultrasound cleaned with organic solvents (ethanol and toluene) for 30', then blown dried with nitrogen. The HF treatment is done in order to create a hydrophobic surface with Si-H groups.

For the brush procedure, the silicon wafer is treated by oxygen plasma in order to generate silanol groups. Immediately after the plasma treatment, the silicon wafer is transferred to closed containers with chlorodimethylsilyl-terminated PS molecules (Polymer Laboratories) diluted in cyclohexane for three hours [38]. After the grafting time was complete, the silicon wafer was removed and ultrasonically cleaned with cyclohexane twice (15' each), in order to eliminate any loosely bound brushes. It was then blown dried with nitrogen and kept in a vacuum oven at 50°C overnight, in order to remove any residual solvents. The $M_W$ ($M_W/M_n$) values have been chosen with respect to the critical $M_W$ around which molecular entanglement occurs ($M_C \approx 30$ kDa) [22]: 7 kDa (1.06), 27.5 kDa (1.06), 135 kDa (1.07). Analysis of the brush homogeneity and thickness has been carried out by means of TOF-SIMS and XPS. The XPS data give a brush thickness of 3 nm for all $M_W$ values. The grafting density ranges from 0,26 to 0,015 $nm^{-2}$ for $M_W$ values from 7,5 to 135 kDa [38]. The $T_g$ of such thin films cannot be measured with our technique. It is however expected that it is higher than $T_g^{bulk}$ if we consider other reports on thicker brushes [43-44] and the phenomenon known as Guiselin brushes [29].

Finally, the PS films are prepared by spin-coating a toluene solution of PS molecules. The thickness is measured by mean of a profilometer (Stylus profiler, Dektak 8). The thermal annealing is performed @ T from 110° to 190°C for time intervals of 10 hours. The PS standards (Polymer Laboratories) employed have following characteristics: 13 kDa (1.03) and 483 kDa (1.05); one above and one below $M_C$.

## 3. Results

At T > $T_g$, polymers are known to have a lower Young's modulus ($E_Y$) and a lower viscosity ($\eta$). In Fig. 1 it is schematically shown that, if T > $T_g$, the tip can penetrate into the film at the application of a non-zero load in a lapse of time ranging from seconds to minutes. This time-dependent deformation of the film can be defined as 'viscous yield'. As we operate at fixed load ($F_N$), the relative distance ($z$) between tip and sample needs to be changed in order to maintain constant cantilever deflection. The graph shown in Fig. 2(a) represents typical indentation ($\delta_V$) versus time ($t$) data obtained for a thick film (> 200 nm, $M_W$ = 483 kDa) at T = 110°C. The film was annealed at a T value ($T_{ann}$) of 130°C. $\delta_V$ is found to be dependent on the contact time ($t$) and the applied load ($F_N$) value.

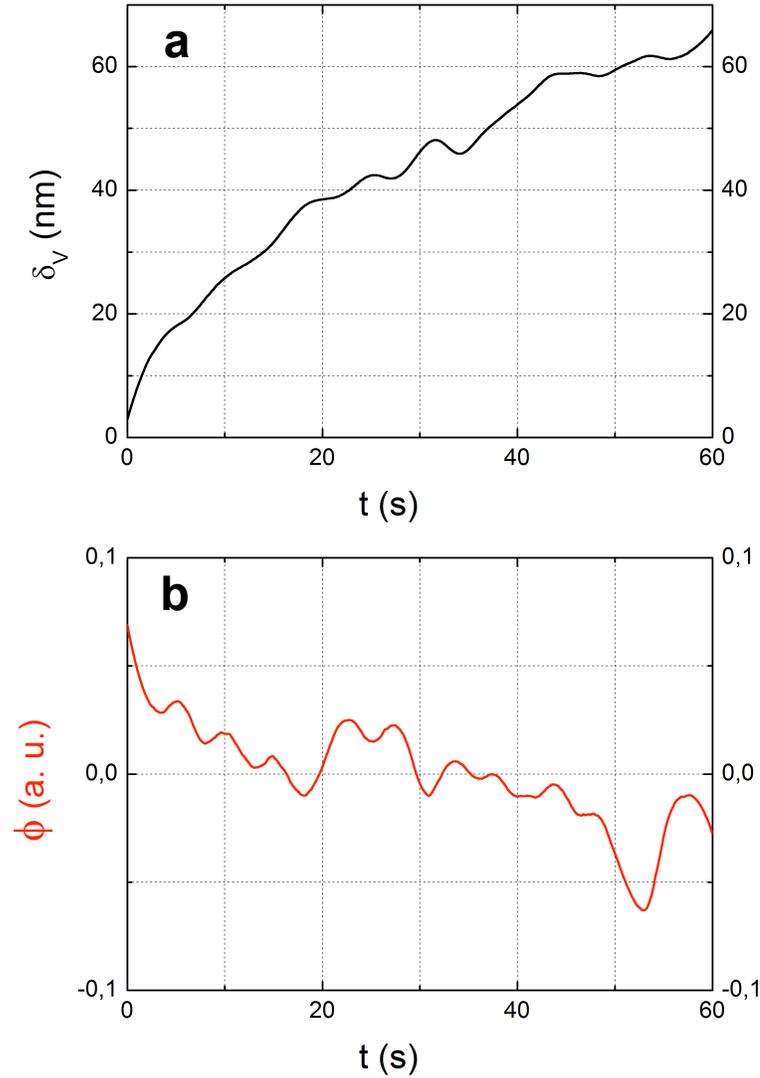

**Fig. 2:** Example of indentation ($\delta_V$) and torsion ($\phi$) versus time ($t$) obtained for a thick (> 200 nm) 483 kDa film at T = 110°C. (a) The insets schematically show that the tip progressively penetrates into the film as T > $T_g$. (b) The insets schematically show that the variations in $\phi$ with time indicate the variations in cantilever torsion during the indentation process.

The cantilever torsion ($\phi$) can also be monitored, telling us if any lateral movement of the tip relative to the surface is occurring during the penetration. In the case reported in Fig. 2(b), it can be noticed that during the indentation $\phi$ varies on two different timescales. The longer timescale shows a variation of the cantilever torsion from an initial non-zero value (due to a tip non perfectly perpendicular to the surface plane) to another value of opposite tilting orientation. This indicates a continuous lateral drift of the tip with respect to the sample, probably due to non-equilibrium temperature of the microscope elements forming the tip/sample loop. On the other hand the shorter timescale indicates more rapid variations of the cantilever torsion, determined by lateral stress accumulated and then released during the indentation process or an artefact due to the cycle time of the heater. When the stress accumulates the tip compresses laterally the PS film that reacts opposing some resistance. This resistance interferes with the indentation making it not a continuous process but with the presence of some bumps.

| | $T_g$ values | |
|---|---|---|
| **Thickness** | **$M_W$ = 483 kDa** | **$M_W$ = 13 kDa** |
| **> 200 nm** | 100-105°C | 90-95°C |
| **≈ 30 nm** | 85-90°C | 80-85°C [a] |
| **Bulk** [17] | ≈ 100°C | ≈ 93°C |

**Table 1.** $T_g$ data as measured by viscous yield for 30 nm and > 200 nm films of 13 kDa and 483 kDa deposited on $SiO_x$ (also reported in Ref. 14). All samples annealed @ 130°C. (a) annealed @ 110°C, otherwise dewetting occurs.

The $T_g$ can be thus evaluated from the observation of the occurrence of viscous yield. In the Tables, we do not indicate a value but a lower and an upper limit, representing the maximum value at which no viscous yield is observed and the minimum value at which viscous yield occurs. We typically perform the measurements with a T interval of 5°C, as they are quite time consuming. This interval is however sufficient for an accurate estimate of $T_g$ variations. In Table 1, we first report the $T_g$ values obtained for PS films deposited on $SiO_x$. It can be noticed that the values for thick films (> 200 nm) are in agreement with the bulk PS at the correspondent $M_W$. For thin films (30 nm), we obtain reduced $T_g$ values, also in agreement with the literature [17,25,39].

In Fig. 3 we schematically report the sample preparations, excluding when we use pristine $SiO_x$. In one case the $SiO_x$ surfaces are treated with a oxygen plasma in order to increase the OH density. Then chlorodimethylsilyl-terminated PS molecules reactive with OH groups are grafted onto them [38]. In such a way PS brushes are prepared with different $M_W$ values, below and above $M_C$ (Fig. 3(c)). In the second case, by means of HF stripping, we obtain H-terminated silicon, which is immediately employed in order to avoid the formation of native oxide (Fig. 3(d)). In both cases a 30 nm PS film is spin-coated on top the substrates (Fig. 3(e) and (f)).

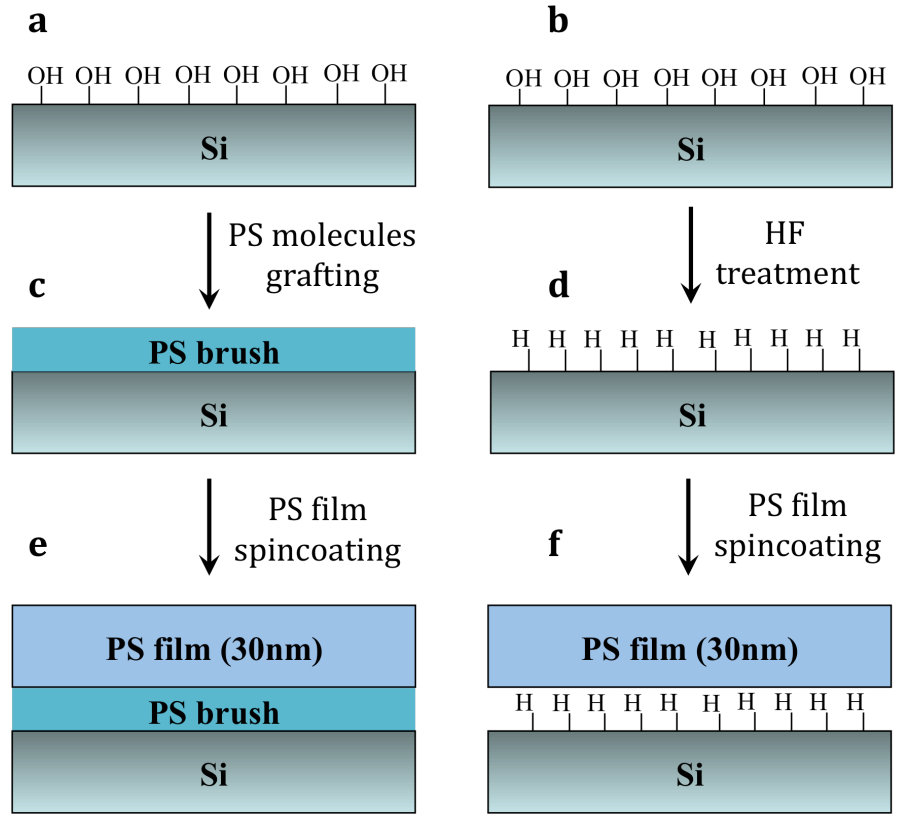

**Fig. 3:** Schematic drawings of the two procedures of sample preparation. (c) PS molecules are chemically attached to $SiO_x$ surfaces by reacting with the Si-OH groups. (e) A film of 30 nm is spin-coated on top of the brush. (d) The native oxide layer is removed by means of HF treatment. (f) A film of 30 nm is spin-coated on top the H-terminated surface.

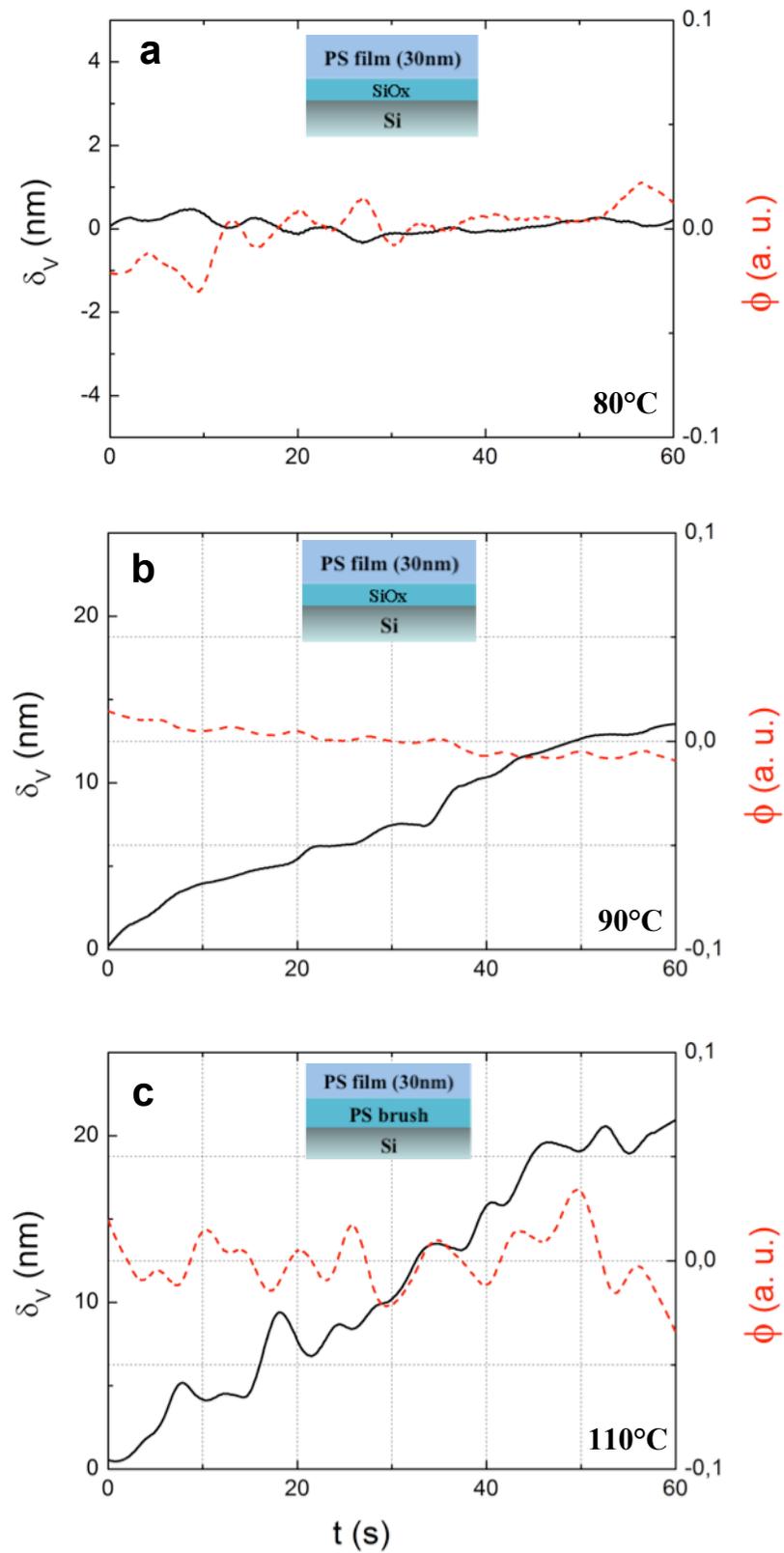

**Fig. 4:** Experimental data for a 30 nm 483 kDa film spin-coated on (a,b) SiO$_x$ and (c) 135 kDa PS brushes. Black continuous lines represent the tip indentation into the film; the red discontinuous lines the cantilever torsion. At T ≤ 85°C viscous yield is negligible (a). It starts to be observed for T ≥ 90°C on $SiO_x$ (b) and for T ≥ 110°C on 135 kDa PS brush (c).

In Fig. 4(a,b) we present typical $\delta_V$ versus $t$ data obtained for a 30 nm 483 kDa film on $SiO_x$ and annealed at $T_{ann}$ = 110°C. For T ≤ 85° C, we observe no viscous deformation taking place with time. For T ≥ 90°C, the viscous deformation appears. In this case the torsion signal does not appreciably vary. This indicates that the lateral drift of the tip with respect to the sample due to non-equilibrium temperature of the microscope elements forming the tip/sample loop is not large.

In Fig. 4(c) we present typical $\delta_V$ versus $t$ data obtained for a 30 nm 483 kDa film deposited on a 135 kDa PS brush and annealed at 160°C. In this case, for T < 110°C, we observe no viscous deformation taking place with time. For T ≥ 110°C, the viscous deformation appears and the torsion signal in the shorter timescale varies much more than in Fig. 4(a): it is more similar to Fig 2(b). This is likely due to the T values at which the measurements have been carried out: at 110°C the thermal lateral drift is larger than at 90°C.

| $M_W = 13\ kDa$ | $T_g$ values | |
|---|---|---|
| **Substrate** | $T_{ann}$ = 130°C | $T_{ann}$ = 160°C |
| **SiO$_X$** | 80-85°C* | - |
| **Si-H** | 80-85°C | 80-85°C |
| **Brush 135 kDa** | 80-85°C | 80-85°C |

\* $T_{ann}$ = 110°C.

**Table 2:** $T_g$ data obtained for 30 nm films with $M_W$ = 13 kDa on PS brushes and H-terminated silicon. On $SiO_x$ $T_{ann}$ needs to be reduced to 110°C otherwise dewetting occurs. The film/brush samples have been annealed in sequence @ 130°C and then @ 160°C, for 10 hours each time.

Thus, in the case of a 30 nm 483 kDa film on a 135 kDa PS brush substrate compared to films of the same thickness on $SiO_x$, the $T_g$ is increased to a value similar to $T_g^{bulk}$. For both measurements, $F_N$ was equal to 50 nN and the tip radius $R_T$ around 50 nm, as measured by means of SEM analysis. The total penetration in both cases is between 15 and 20 nm. Considering the elastic deformation induced at the moment of contact creation, this value is compatible with the film thickness [14].

In Tables 2 and 3, we summarize the $T_g$ values obtained for 13 and 483 kDa thin films, respectively, deposited on H-terminated silicon and PS brushes. A first notation needs to be explained with regards to thermal annealing. On PS brushes, independently of the $M_W$ of the brush and of the film, annealing can be performed at $T_{ann}$ up to 190°C for a rather long time (10 hours). No dewetting phenomenon has been ever observed, whereas it can be easily obtained for $T_{ann} = 130$°C on such thin films directly spun on top of $SiO_x$. This is due to the improved compatibility between the surface and the film that induces a good wettability [40].

| $M_W = 483\ kDa$ | $T_g$ values | |
|---|---|---|
| **Substrate** | $T_{ann} = 130$°C | $T_{ann} = 160$°C |
| **$SiO_X$** | 85-90°C* | - |
| **Si-H** | 100-105°C | 100-105°C |
| **Brush 7,5 kDa** | 80-85°C | 80-85°C |
| **Brush 27.5 kDa** | 100-105°C | 100-105°C |
| **Brush 135 kDa** | 95-100°C | 105-110°C |

\* $T_{ann} = 110$°C.

**Table 3:** $T_g$ data obtained for 30 nm films with $M_W = 483$ kDa on PS brushes and H-terminated silicon. On $SiO_x$ $T_{ann}$ needs to be reduced to 110°C otherwise dewetting occurs. The film/brush samples have been annealed in sequence @ 130°C and then @ 160°C, for 10 hours each time.

Starting with Table 2, we notice that 13 kDa films present the same $T_g$ value, independently of the substrate. Much more interestingly the situation changes for 483 kDa thin films, as reported in Table 3: on H-treated silicon surface the $T_g$ value increases back to $T_g^{bulk}$. The behaviour of thin films on the PS brushes depends on the $M_W$ values. For $M_W < M_C$, no variation can be observed. For $M_W \geq M_C$, the $T_g$ values become equal to and higher than $T_g^{bulk}$, such as in the case of the 135 kDa brush annealed at 160°C. Notably all film/brush samples have undergone the same thermal treatment.

## 4. Discussion

Our observations can be generalized as follows. The $T_g$ of thin films on $SiO_x$ decreases with decreasing thickness independently of the $M_W$, as reported elsewhere [5-6,41]. When the brush/film systems are both made of molecules with $M_W > M_C$ the $T_g$ of thin films increases with respect to the values on $SiO_x$. In the case of 135 kDa brush, $T_g$ also depends on $T_{ann}$. In particular if $T_{ann} = 130$°C the $T_g$ is slightly lower than $T_g^{bulk}$. If $T_{ann} = 160$°C the $T_g$ becomes slightly higher than $T_g^{bulk}$. However this difference from $T_g^{bulk}$ is within the T interval between two successive measurements and it is probably not significant.

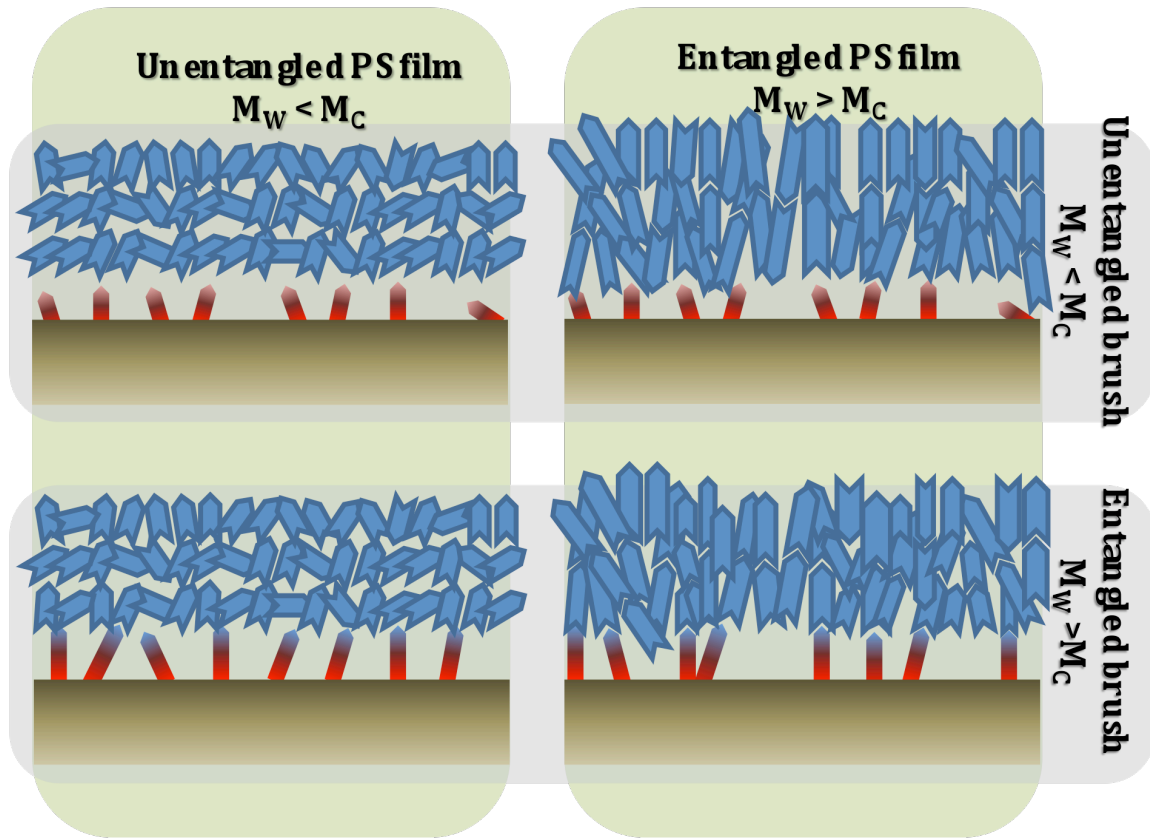

**Fig. 5:** Schematic drawing of different film/brush systems. Each segment represents a molecule ($M_W < M_C$) or a portion of a molecule between two entanglements ($M_W > M_C$). (a, c) A 13 kDa film on top of a PS brush. Entanglement can never occur as the film has $M_W < M_C$. (b, d) A 483 kDa film on a PS brush. If the brush has $M_W > M_C$, then the molecules can entangle In (d), the entanglements are indicated by the interconnection of arrows. If the brush has $M_W < M_C$, no entanglement can occur.

Our interpretation of these results is based on the occurrence or not of molecular entanglement between molecules present in the film and in the brush. To our knowledge a direct measurement of the entanglement occurrence is not possible at the moment, especially for very thin films. Therefore one needs to infer it from indirect evidence such as the present case, where the $T_g$ of thin films increases back to the $T_g^{bulk}$ when both film and brush have a $M_W > M_C$.

It is not possible to draw a realistic picture of the topological disposition of the molecules and also to describe what occurs at the brush/film interface. In Fig. 5 we report a schematic view of the different PS film/brush systems that can be useful for the following discussion. When one of the two components (film or brush) has a $M_W < M_C$, no molecular entanglement can occur (see case (a), (b) and (c)). The dynamics of the film molecules is not affected and only the wettability is improved. When both the two components (film and brush) have a $M_W > M_C$, the molecules at the interface can interpenetrate and form a wide superposed region [42]. In this region molecular entanglement can occur (case (d)) and the dynamics of the film molecules can be affected, i.e. $T_g \geq T_g^{bulk}$. The sample can be considered equivalent to the case of a 30 nm thick PS film placed onto a thick PS film with the same $M_W$.

In the literature one can find a number of studies interested in the properties of polymer brushes, for instance adhesive [42] or mechanical ones [43]. However, if one restricts the research to papers related to our investigation, three papers have reported data on related experiments. Although at a first sight it is not apparent, we believe their results are compatible with our findings. In particular, Lee et al. [39] have carried out systematic experiments of a film with $M_W > M_C$ on brushes with $M_W$ from below to slightly above $M_C$. They clearly state that their interest lies in the role played by the interfacial energy. This is the reason why they employ PS brushes of various $M_W$. In particular, for a 33 nm thick film with $M_W = 100$ kDa (1.02), they report an increase in $T_g$ when placed on a PS brush with $M_W = 38$ kDa compared to PS brushes with $M_W < M_C$. However this increase is not enough for Tg to be equal to $T_g^{bulk}$. It is important to notice that they have annealed the samples at 120°C. Although they never mention the molecular entanglement, it is to be expected to occur when $T_{ann}$ is higher than the $T_g$ of both film and brush. According to the literature, the brush $T_g$ can be much higher than $T_g^{bulk}$. This is due to the fact that the molecules are bound to the substrate and therefore the dynamics can be strongly affected [44-45]. In our case unfortunately the $T_g$ of such thin brushes cannot be measured with our technique. We can however infer from Table 3, that the $T_g$ of a 27 kDa brush might be inferior to 130°C, whereas the $T_g$ of a 135 kDa brush might be higher. From these data we can then deduce that the $T_g$ of a brush with $M_W = 38$ kDa should be close to 120°C, although this might depend on the grafting density. This could explain why they obtain a $T_g$ higher than for $M_W < M_C$ but still not equal to $T_g^{bulk}$: molecular entanglements have not been fully established. Compared to our experiments, this situation can be compared to that one of a 483 kDa film on a 135 kDa brush and annealed at 130°C.

The second relevant paper is by Tsui et al. [17]. In Fig. 3 of their manuscript, it can be noticed that the $T_g$ of a 33 nm thick PS film with $M_W = 96$ kDa (1.02) increases when there is a PS brush with $M_W = 10$ kDa ($M_W/M_n = 1.1$ to 1.8). However $T_g$ is still not equal to $T_g^{bulk}$. The authors suggest that this increase might be due to the fact that brush molecules fully penetrate into the film (as proven by neutron scattering measurements) making it thicker. Our experimental data suggest that the increase may be due to the occurrence of molecular entanglements. Although the $M_W$ of the brush is still lower than $M_C$, their $M_w$ value is higher than our lower value and the $M_W/M_n$ is quite large. Therefore, molecular entanglement might still partially occur. Unfortunately, the authors do not report the value of $T_{ann}$.

Finally, the third relevant paper by Clough et al. reports data in good agreement with our results [46]. In particular they show that the $T_g$ of thin PS films with $M_W = 41$ kDa on a 96 kDa brush does not change from the $T_g^{bulk}$. For completion, they also observe that the $T_g$ of thin PS films with $M_W = 41$ kDa on a 10 kDa brush is slightly higher than on $SiO_x$. This last observation is partially not in agreement with what we observe. However small this discrepancy might be, it may suggest that an additional effect affecting the $T_g$ of thin films and not imputable to molecular entanglement could be present.

In Fig. 6, we sketch the different systems made of films with $M_W$ below and above $M_C$ and the H-terminated silicon surface. Our interpretation of these results is the following. Where the molecules are in direct contact with the surface, single portions of the molecules themselves can be immobilized due to the strong interaction with the substrate. This determines slower dynamics of the molecules themselves. If the two free strands of the molecule branching out from those bonds are entangled with the molecules of the film, the sample can be considered equivalent to the sample in Fig. 5(d). If not, the sample is equivalent to the sample in Fig. 5(a). Annealing at 130°C is sufficient to establish conditions that determine the maximum $T_g$ value. These data indicate that a strong interfacial interaction of the PS molecules with the substrate cannot affect the film properties unless molecular entanglement is present.

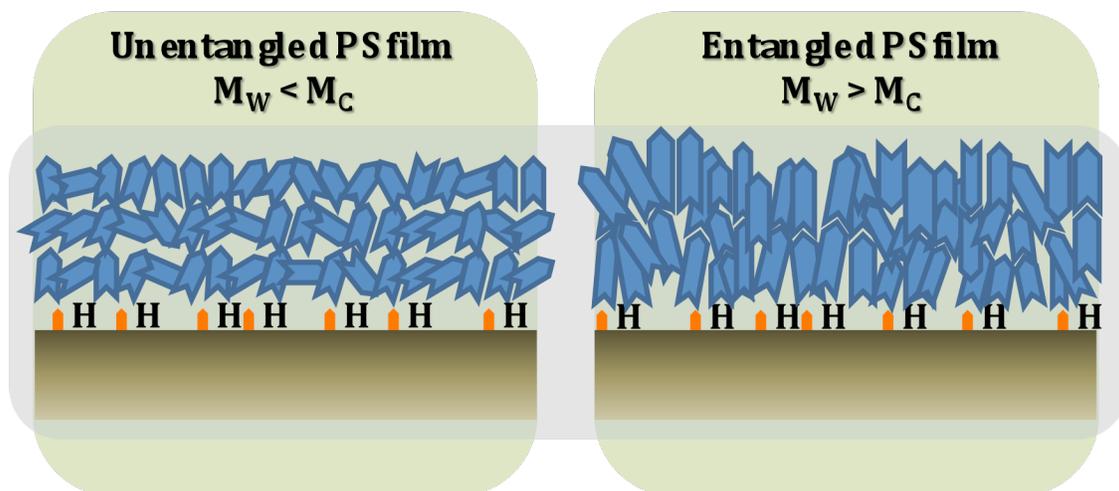

**Fig. 6:** Schematic drawing of different film/substrate systems. (a) 13 kDa and (b) 483 kDa films on H-terminated silicon surfaces. The portions of the molecules in direct contact with the substrate are blocked. The molecules of the 483 kDa film are entangled with the molecules of the film, the molecules of the 13 kDa film are not.

## 5. Conclusions

We have shown that the glass transition temperature of films made of polystyrene molecules with molecular weight below the critical value for the occurrence of molecular entanglement cannot be affected neither by the hydrogen-terminated silicon nor by the presence of polystyrene brushes. On the contrary the glass transition temperature of films made of polystyrene molecules with molecular weight above this critical value is equal to the bulk one, and sometimes even higher, both on the hydrogen-terminated silicon and on the brushes made of polystyrene molecules with molecular weight above the critical value.

Based on these findings, we can state that a strong interaction with the substrate is relevant in order to increase the glass transition temperature of thin polystyrene films. However, this effect is subordinated to the presence of molecular entanglement. It has thus been demonstrated that, in order to control the viscoelastic properties of thin polystyrene films versus temperature, molecular entanglement is highly relevant. This effect is potentially important for technological applications where the glass transition of thin films needs to be varied and controlled, as exampled in the case of nanoimprint lithography (NIL).

### Acknowledgements

The Danish Research Council, The Aarhus Graduate School of Science (AGoS) and the iNANO Centre are thanked for funding for MP.